\documentstyle[12pt,epsfig]{article}
\def\lbldef#1#2{\expandafter\gdef\csname #1\endcsname {#2}}

\def\href#1#2{#2}  

\begin{document}
\baselineskip=15.5pt
\pagestyle{plain}
\setcounter{page}{1}

\begin{titlepage}

\begin{flushright}
CERN-TH/99-295\\
hep-th/9909180
\end{flushright}
\vspace{10 mm}

\begin{center}
{\Large  (Generalized) Conformal Quantum Mechanics of 0-Branes and 
Two-Dimensional Dilaton Gravity}

\vspace{5mm}

\end{center}

\vspace{5 mm}

\begin{center}
{\large Donam Youm\footnote{Donam.Youm@cern.ch}}

\vspace{3mm}

Theory Division, CERN, CH-1211, Geneva 23, Switzerland

\end{center}

\vspace{1cm}

\begin{center}
{\large Abstract}
\end{center}

\noindent

We study the relation between the (generalized) conformal quantum 
mechanics of 0-branes and the two-dimensional dilaton gravity.  The 
two-dimensional actions obtained from the supergravity effective 
actions for the (dilatonic) 0-branes through the compactification on a 
sphere are related to known two-dimensional dilaton gravity models.  
The $SL(2,{\bf R})$ symmetry of the (generalized) conformal quantum 
mechanics is realized within such two-dimensional models.  The 
two-dimensional dilatonic gravity model derived from the non-dilatonic 
0-brane action is related to the Liouville theory and therefore is 
conformal, whereas the two-dimensional model derived from the dilatonic 
0-brane action does not have the conformal symmetry.  

\vspace{1cm}
\begin{flushleft}
CERN-TH/99-295\\
September, 1999
\end{flushleft}
\end{titlepage}
\newpage

\section{Introduction}

According to the AdS/CFT duality conjecture \cite{mal1} and its 
generalization \cite{mal2}, the bulk (gravity) theory on the near-horizon 
manifold of the supergravity brane solution is equivalent to the boundary 
(field) theory, which is the worldvolume theory of the brane in the 
decoupling limit.  This conjecture relates closed string theory and the open 
string theory in the appropriate limit.  Namely, in taking the decoupling 
limit of the brane worldvolume theory, which is the open string theory on 
the brane, to obtain the gauge field theory (an extensive review on the 
gauge field theory as the decoupling limit of the brane worldvolume theory 
can be found in Ref. \cite{giv}), one ends up with the near horizon 
region \cite{gib1,gib2} of the brane supergravity solution, which is the 
low energy limit of the closed string theory.  This AdS/CFT duality 
conjecture can be regarded as a concrete string theory realization of the 
holographic principle \cite{tho,sus1,sus2} and the previously conjectured 
equivalence \cite{duf1,gib1} between the bulk theory on the AdS space and 
the supersingleton field theory, which is the effective worldvolume theory 
of the brane solutions, on the boundary of the AdS space (for a review on 
this subject, see Ref. \cite{duf2}).   

The isometry symmetry of the near-horizon manifold manifests as a symmetry 
of the boundary field theory and the boundary field theory is conformal when 
the near-horizon geometry of the supergravity brane solution (in the 
string-frame) contains the AdS space \cite{gib1,gib2}.   On the other hand, 
according to Refs. \cite{duf3,gib1,gib2,tow1}, any $p$-brane supergravity 
solutions in the near-horizon limit take the AdS$\times S^n$ form in the 
so-called dual-frame.  So, the isometry symmetry of the AdS space in the 
near-horizon region of the dilatonic brane solutions in the dual-frame is also 
expected to be present in the corresponding boundary field theories but the 
boundary theories are not genuinely conformal due to non-trivial dilaton 
field.  Nevertheless, one can still define generalized conformal field 
theory \cite{jev1,jev2}, where string coupling is now regarded as a part 
of background fields that transform under the symmetry.  

In the 0-brane case, the corresponding boundary theory can be thought of as 
the (generalized) conformal quantum mechanics 
\cite{aff,ap,fr,kal,tow2,you,bbh} of the probe 0-brane in the near-horizon 
background of the source 0-brane supergravity solution.  The (generalized) 
conformal quantum mechanics of the 0-brane is shown \cite{kal,you,bbh} to 
have the $SL(2,{\bf R})$ symmetry, as expected from the fact that the 
near-horizon limit of the 0-brane solution (in the dual-frame) is of the 
form AdS$_2\times S^n$, where the isometry of the AdS$_2$ part is $SO(1,2)
\cong SL(2,{\bf R})$.  The $SL(2,{\bf R})$ symmetry of the probe 0-brane is 
conformal for the non-dilatonic 0-branes \cite{kal}, and is not genuinely 
conformal but can be extended to the generalized conformal symmetry for the 
dilatonic 0-branes \cite{you,bbh}.  

Since the supergravity 0-brane solution (in the dual-frame) in the 
near-horizon region takes AdS$_2\times S^n$ form, one can perform the 
Freund-Rubin compactification \cite{fre} of the 0-brane supergravity action 
on $S^n$.  The resulting theory is the two-dimensional dilaton gravity with 
the (dilaton dependent) cosmological constant term.  Therefore, one would 
expect the relevance of the (generalized) conformal quantum mechanics of 
0-branes to the two-dimensional dilaton gravity.  It is the purpose of this 
paper to elaborate on such relation.  We observe that the two-dimensional 
dilaton gravity compactified from the 0-brane effective action has 
the $SL(2,{\bf R})$ affine symmetry, in accordance with the fact that 
the (generalized) conformal quantum mechanics of 0-brane has the 
$SL(2,{\bf R})$ symmetry.  Also, consistently with the case of the 
(generalized) conformal quantum mechanics of 0-branes, the two-dimensional 
gravity model obtained from the non-dilatonic 0-brane action is conformal, 
whereas the one obtained from the dilatonic 0-brane action is not conformal.

The paper is organized as follows.  In section 2, we summarize the general 
supergravity 0-brane solutions in $D$ dimensions.  In section 3, we summarize 
the (generalized) conformal quantum mechanics of 0-branes.  In section 4, we 
compactify the supergravity actions for the 0-branes on $S^{D-2}$ and relate 
the resulting two-dimensional actions to various two-dimensional dilaton 
gravity models for the purpose of studying the relevance of the (generalized) 
conformal quantum mechanics of 0-branes to the two-dimensional dilaton 
gravity theories.

\section{Supergravity 0-Brane Solutions}

In this section, we summarize the general supergravity 0-brane solution 
with an arbitrary dilaton coupling parameter $a$ in arbitrary spacetime 
dimensions $D$.  The supergravity action in the Einstein-frame for such 
dilatonic 0-brane is given by
\begin{equation}
S_E={1\over{2\kappa^2_D}}\int d^Dx\sqrt{-G^E}\left[{\cal R}_{G^E}
-{4\over{D-2}}(\partial\phi)^2-{1\over{2\cdot 2!}}e^{2a\phi}F^2_2\right],
\label{ddim0act}
\end{equation}
where $\kappa_D$ is the $D$-dimensional Planck constant, $\phi$ is the 
$D$-dimensional dilaton, and $F_2=dA_1$ is the field strength of the 1-form 
potential $A_1=A_{M}dx^M$ ($M=0,1,...,D-1$).  The extreme 0-brane solution 
to the equations of motion of this action is given by
\begin{eqnarray}
ds^2_E&=&-H^{-{{4(D-3)}\over{(D-2)\Delta}}}dt^2+H^{4\over{(D-2)\Delta}}
(dr^2+r^2d\Omega^2_{D-2}),
\cr
e^{\phi}&=&H^{{(D-2)a}\over{2\Delta}},\ \ \ \ 
A_t=-H^{-1},
\label{0branesol}
\end{eqnarray}
where $H=1+\left({{\mu}\over{r}}\right)^{D-3}$ and $\Delta={{(D-2)a^2}\over
{2}}+{{2(D-3)}\over{D-2}}$.  The examples on the values of $\Delta$ and $a$ 
for some interesting cases are
\begin{itemize}
\item D0-brane in $D=10$: ($a$,$\Delta$)=(${3\over 4}$,4)
\item d0-brane in $D=6$: ($a$,$\Delta$)=(${1\over 2}$,2)
\item dilatonic black hole in $D=4$: ($a$,$\Delta$)=($\sqrt{3}$,4), (1,2), 
(${1\over\sqrt{3}}$,${4\over 3}$), (0,1)
\end{itemize}

One can perform the Weyl rescaling transformation $G^E_{MN}=e^{-{4\over 
{D-2}}\phi}G^{s}_{MN}$ to obtain the following ``string-frame'' effective 
action for the 0-brane in $D$ dimensions:
\begin{equation}
S_{s}={1\over{2\kappa^2_D}}\int d^Dx\sqrt{-G^{s}}\left[e^{-2\phi}
\left\{{\cal R}_{G^{s}}+4(\partial\phi)^2\right\}-{1\over{2\cdot 2!}}
e^{2(a-{{D-4}\over{D-2}})\phi}F^2_2\right].
\label{0branactstr}
\end{equation}
In the special cases pointed out in the previous paragraph, i.e. the cases 
with the dilaton coupling $a=(D-4)/(D-2)$, this action takes the form of 
the string-frame effective actions of the string theory.  The spacetime 
metric for the dilatonic 0-brane in this frame is given by
\begin{equation}
ds^2_{s}=-H^{-{{4(D-3)-2a(D-2)}\over{(D-2)\Delta}}}dt^2
+H^{{4+2a(D-2)}\over{(D-2)\Delta}}(dr^2+r^2d\Omega^2_{D-2}).
\label{0branesolstr}
\end{equation}
Note, in the case of $D=10$ and ($a$,$\Delta$)=(${3\over 4}$,4), 
one recovers the following familiar string-frame D0-brane solution:
\begin{eqnarray}
ds^2_{s}&=&-H^{-{1\over 2}}dt^2+H^{1\over 2}(dr^2+r^2d\Omega^2_8),
\cr
e^{\phi}&=&H^{3\over 4},\ \ \ \  A_t=-H^{-1};\ \ \ \ \ \  
H=1+\left({{\mu}\over{r}}\right)^7.
\label{d0bransolstr}
\end{eqnarray}
Also, in the case of $D=6$ and ($a$,$\Delta$)=(${1\over 2},2$), the 
solution (\ref{0branesolstr}) reduces to the following string-frame 
solution for d0-brane in $D=6$:
\begin{eqnarray}
ds^2_{s}&=&-H^{-1}dt^2+H(dr^2+r^2d\Omega^2_4),
\cr
e^{\phi}&=&H^{1\over 2},\ \ \ \ A_t=-H^{-1};\ \ \ \ \ \  
H=1+\left({{\mu}\over{r}}\right)^3.
\label{6dim0bransolstr}
\end{eqnarray}

One can also think of the dilatonic 0-brane in $D$ dimensions as being 
magnetically charged under the $(D-2)$-form field strength $F_{D-2}$.  
The corresponding Einstein-frame effective action is
\begin{equation}
\tilde{S}_E={1\over{2\kappa^2_D}}\int d^Dx\sqrt{-G^E}\left[{\cal R}_{G^E}
-{4\over{D-2}}(\partial\phi)^2-{1\over{2\cdot (D-2)!}}e^{-2a\phi}F^2_{D-2}
\right],
\label{ddim0actdual}
\end{equation}
where this action is obtained from the action (\ref{ddim0act}) through the 
electric-magnetic duality transformation.  So, the dilatonic 0-brane solution 
(\ref{0branesol}) takes the following form in terms of the ``dual'' field 
parametrization:
\begin{eqnarray}
ds^2_E&=&-H^{-{{4(D-3)}\over{(D-2)\Delta}}}dt^2+H^{4\over{(D-2)\Delta}}
(dr^2+r^2d\Omega^2_{D-2}),
\cr
e^{\phi}&=&H^{{(D-2)a}\over{2\Delta}},\ \ \ \ 
F_{D-2}=\star(dH\wedge dt).
\label{0branesoldual}
\end{eqnarray}
The following effective action in the dual-frame \cite{bbh}, in which all 
the 0-brane solutions in the near-horizon region take the AdS$_s\times 
S^{D-2}$ form, is obtained through the Weyl rescaling transformation 
$G^E_{MN}=e^{-{2a\over{D-3}}\phi}G^{d}_{MN}$:
\begin{equation}
\tilde{S}_{d}={1\over{2\kappa^2_D}}\int d^Dx\sqrt{-G^{d}}e^{\delta\phi}
\left[{\cal R}_{G^{d}}+\gamma(\partial\phi)^2-{1\over{2\cdot (D-2)!}}
F^2_{D-2}\right],
\label{ddim0actdual2}
\end{equation}
where
\begin{equation}
\delta\equiv -{{D-2}\over{D-3}}a,\ \ \ \ \ 
\gamma\equiv {{D-1}\over{D-2}}\delta^2-{4\over{D-2}}.
\label{defdelgam}
\end{equation}
In the dual-frame, the dilatonic 0-brane solution takes the following 
form:
\begin{equation}
ds^2_d=-H^{{2\Delta-4(D-3)}\over{(D-3)\Delta}}dt^2+H^{2\over{D-3}}
(dr^2+r^2d\Omega^2_{D-2}),
\label{0branesoldfram}
\end{equation}
where the dilaton and the $(D-2)$-form field strength take the same form as 
in Eq. (\ref{0branesoldual}).  In the near horizon region ($r\ll\mu$), which 
corresponds to the decoupling limit of the corresponding boundary worldvolume 
theory, the metric (\ref{0branesoldfram}) takes the following 
AdS$_2\times S^{D-2}$ form \cite{bbh}:
\begin{equation}
ds^2_d\approx -\left({{\mu}\over r}\right)^{2-{{4(D-3)}\over{\Delta}}}dt^2
+\left({{\mu}\over{r}}\right)^2dr^2+\mu^2d\Omega^2_{D-2},
\label{nearhordualmet}
\end{equation}
and when the 0-brane is regarded as being electrically charged under the 
1-form potential, which is relevant for the (generalized) conformal quantum 
mechanics of the probe 0-brane, the metric takes the same form as the above 
and the dilaton and the non-zero component of the $U(1)$ gauge field are
\begin{equation}
e^{\phi}\approx \left({{\mu}\over{r}}\right)^{{(D-2)(D-3)a}\over{2\Delta}},
\ \ \ \ \ 
A_t\approx-\left({r\over\mu}\right)^{D-3}.
\label{nearhordualmet2}
\end{equation}
One can bring this AdS$_2\times S^{D-2}$ metric to the following 
standard form in the horospherical coordinates by introducing new radial 
coordinate $\bar{r}={\Delta\over{2(D-3)-\Delta}}\mu^{{2\Delta-2(D-3)}
\over\Delta}r^{{2(D-3)-\Delta}\over\Delta}$:
\begin{equation}
ds^2_d\approx -\left({{\bar{r}}\over{\bar{\mu}}}\right)^2dt^2+
\left({{\bar{\mu}}\over{\bar{r}}}\right)^2d\bar{r}^2+
\mu^2\Omega^2_{D-2},
\label{horomet}
\end{equation}
and when the 0-brane is regarded as being electrically charged under the 
1-form potential the metric takes the same form as the above and the dilaton 
and the non-zero component of the $U(1)$ gauge field are
\begin{equation}
e^{\phi}\approx\left({\bar{\mu}\over\bar{r}}\right)^{{(D-2)(D-3)a}
\over{4(D-3)-2\Delta}},\ \ \ \ \ 
A_t=-\left({\bar{r}\over\bar{\mu}}\right)^{{(D-3)\Delta}\over{2(D-3)-\Delta}},
\label{horomet2}
\end{equation}
where $\bar{\mu}\equiv{{\Delta\mu}\over{2(D-3)-\Delta}}$.

\section{(Generalized) Conformal Quantum Mechanics of 0-Branes}

The boundary theory counterpart to the bulk gravity theory in the near-horizon 
background of (dilatonic) 0-branes can be thought of as the (generalized) 
conformal quantum mechanics.  The (generalized) conformal mechanics is 
described by the dynamics of the probe (dilatonic) 0-brane moving in the 
near-horizon field background (\ref{nearhordualmet}) or (\ref{horomet}) of 
the stack of the large number of source (dilatonic) 0-branes.  The action 
for the probe (dilatonic) 0-brane with the mass $m$ and the charge $q$ is
\begin{equation}
S_{\rm probe}=\int d\tau {\cal L}=\int d\tau
\left(me^{-{{D-4}\over{D-3}}\phi}\sqrt{-G^d_{MN}\dot{x}^M\dot{x}^N}
-q\dot{x}^MA_M\right),
\label{probact}
\end{equation}
where $G^d_{MN}$, $A_M$ and $\phi$ are the dual-frame metric, the $U(1)$ 
gauge field and the dilaton field produced by the source (dilatonic) 
0-brane.  

From the following mass-shell constraint for the probe (dilatonic) 0-brane:
\begin{equation}
G^{d\,MN}(P_M-qA_M)(P_N-qA_N)+m^2e^{-2{{D-4}\over{D-3}}\phi}=0,
\label{masshell}
\end{equation}
where $P_M=-\delta{\cal L}/\delta\dot{x}^M$ is the canonical momentum, 
one obtains the expression for the Hamiltonian ${\cal H}=-P_t$ describing the 
mechanics of the probe 0-brane.  With the following general spherically 
symmetric Ansatz for the spacetime metric (in the dual-frame):
\begin{equation}
G^d_{MN}dx^Mdx^N=-A(r)dt^2+B(r)dr^2+C(r)d\Omega^2_{D-2},
\label{sphsymmet}
\end{equation}
one obtains the following expression for the Hamiltonian \cite{kal,you,bbh}:
\begin{equation}
{\cal H}={{P^2_r}\over{2f}}+{{g}\over{2f}},
\label{hamilton}
\end{equation}
where $f$ and $g$ are given by
\begin{eqnarray}
f&\equiv&{1\over 2}A^{-{1\over 2}}Be^{-{{D-4}\over{D-3}}\phi}\left[
\sqrt{m^2+e^{2{{D-4}\over{D-3}}\phi}(P^2_r+BC^{-1}\vec{L}^2)/B}
+qA^{-{1\over 2}}A_te^{{{D-4}\over{D-3}}\phi}\right],
\cr
g&\equiv&Be^{-2{{D-4}\over{D-3}}\phi}\left[(m^2-q^2A^{-1}A^2_t
e^{2{{D-4}\over{D-3}}\phi})+C^{-1}e^{2{{D-4}\over{D-3}}\phi}
\vec{L}^2\right],
\label{fgdef}
\end{eqnarray}
where $\vec{L}^2$ is the angular momentum operator of the probe 0-brane, 
and the expressions for $A$, $B$, $C$, $\phi$ and $A_t$ that should enter 
in these expressions for $f$ and $g$ can be read off from the near 
horizon 0-brane solution (\ref{nearhordualmet}) with (\ref{nearhordualmet2}) 
or (\ref{horomet}) with (\ref{horomet2}).  For particular values of the 
dilaton coupling parameter given by $a=(D-4)/(D-2)$
\footnote{For this particular value of the dilatonic coupling parameter $a$, 
the kinetic term for the $U(1)$ gauge field $A_1$ in the ``string-frame'' 
(Cf. see the corresponding action (\ref{0branactstr}) in the 
``string-frame'') does not have the dilaton $\phi$ dependence, which is 
also the characteristic of the kinetic terms for the RR form fields of 
string theories.}, 
which include the cases of D0-brane, d0-brane and $D=4$ Reissner-Nordstr\"om 
black hole, the expressions for $f$ and $g$ get simplified significantly 
\cite{you,bbh} as follows:
\begin{eqnarray}
f&=&{1\over 2}A^{-{1\over 2}}Be^{-{{D-4}\over{D-3}}\phi}\left[
\sqrt{m^2+e^{2{{D-4}\over{D-3}}\phi}(P^2_r+BC^{-1}\vec{L}^2)/B}+q\right],
\cr
g&=&Be^{-2{{D-4}\over{D-3}}\phi}\left[(m^2-q^2)+C^{-1}
e^{2{{D-4}\over{D-3}}\phi}
\vec{L}^2\right].
\label{simpfg}
\end{eqnarray}
In this case, in the extreme limit ($m=q$) of the probe 0-brane the first 
term in $g$ drops out and when the (extreme) probe's motion is restricted 
along the radial direction (i.e. $\vec{L}^2=0$), $g=0$.  

The mechanics of the probe 0-brane has the $SL(2,{\bf R})$ symmetry with the 
following generators:
\begin{equation}
{\cal H}={{P^2_r}\over{2f}}+{{g}\over{2f^2}},\ \ \ \ 
{\cal K}=-{1\over 2}fr^2,\ \ \ \ 
{\cal D}={1\over 2}rPr,
\label{sl2rgens}
\end{equation}
where the Hamiltonian ${\cal H}$ generates the time translation, 
${\cal K}$ generates the special conformal transformation and ${\cal D}$ 
generates the scale transformation or the dilatation.  These generators 
satisfy the following $SL(2,{\bf R})$ algebra \cite{kal,you,bbh}:
\begin{equation}
[{\cal D},{\cal H}]={\cal H},\ \ \ \ 
[{\cal D},{\cal K}]=-{\cal K},\ \ \ \ 
[{\cal H},{\cal K}]=2{\cal D}.
\label{sl2ralg}
\end{equation}

In the non-dilatonic case ($a=0$), i.e. the case of the Reissner-Nordstr\"om 
solution, this $SL(2,{\bf R})$ symmetry is the genuine conformal symmetry 
\cite{kal}.  But when the dilaton field is non-trivial ($a\neq 0$), the above 
$SL(2,{\bf R})$ symmetry is no longer conformal, because the string coupling 
$g_s=e^{\phi_{\infty}}$ changes under the dilatation and the special 
conformal transformation.  But one can still think of the ``generalized'' 
$SL(2,{\bf R})$ conformal symmetry of the quantum mechanics of the dilatonic 
0-brane \cite{you,bbh}, in which the string coupling is now regarded as a 
part of background fields that transform under this symmetry 
\cite{jev1,jev2}.

\section{Relations to Two-Dimensional Dilaton Gravity}

In this section, we elaborate on the relation of the (generalized) conformal 
quantum mechanics of the (probe) 0-branes to the two-dimensional dilaton 
gravity theories.  We begin by discussing the symmetries of the general 
class of two-dimensional dilaton gravity theories, to which we shall 
relate the $S^{D-2}$-compactified effective supergravity action for the 
(source) 0-branes.  

The most general coordinate invariant action functional of the metric 
$g_{\mu\nu}$ and the dilaton $\phi$, which depends at most on two 
derivatives of the fields, in two spacetime dimensions has the 
following form \cite{ban}:
\begin{equation}
S[g,\phi]={1\over{2\kappa^2_2}}\int d^2x\sqrt{-g}\left[{1\over 2}
\partial_{\mu}\phi\partial^{\mu}\phi+V(\phi)+D(\phi){\cal R}\right],
\label{gen2dact}
\end{equation}
where $V(\phi)$ is an arbitrary function of $\phi$, $D(\phi)$ 
is a differentiable function of $\phi$ such that $D(\phi)\neq 0$ 
and $dD(\phi)/d\phi\neq 0$, and ${\cal R}$ is the Ricci scalar 
of the metric $g_{\mu\nu}$.  As pointed out in Ref. \cite{ban}, 
although the action (\ref{gen2dact}) is expressed in terms of 
two functions $V(\phi)$ and $D(\phi)$ of $\phi$, the physics of this model 
does not depend on two arbitrary functions of $\phi$.  In fact, 
it is later explicitly shown (following a comment in Ref. \cite{ban}) in 
Ref. \cite{mgk} that a $\phi$-dependent Weyl rescaling transformation 
$\bar{g}_{\mu\nu}=\Omega^2(\phi)g_{\mu\nu}$, where $\Omega(\phi)$ is 
the solution to the following differential equation:
\begin{equation}
{1\over 2}-2{{dD}\over{d\phi}}{{d\ln\Omega}\over{d\phi}}=0,
\label{diffeqweyl}
\end{equation}
followed by redefinition of the dilaton field $\bar{\phi}\equiv D(\phi)$ 
leads to the following action expressed in terms of only one function 
of the dilaton:
\begin{equation}
S={1\over{2\kappa^2_2}}\int 
d^2x\sqrt{-\bar{g}}\left[\bar{\phi}{\cal R}_{\bar{g}}
+\bar{V}(\bar{\phi})\right],
\label{simp2dact}
\end{equation}
where the potential $\bar{V}$ is given by
\begin{equation}
\bar{V}(\bar{\phi})={{V(\phi(\bar{\phi}))}\over{\Omega^2(\phi(\bar{\phi}))}}.
\label{2dpot}
\end{equation}
From now on, we suppress the bars in the fields.

In the conformal coordinates, in which the metric takes the form $g_{\mu\nu}
=e^{2\rho}\eta_{\mu\nu}$, the general action (\ref{simp2dact}) can be put 
into the following form of the non-linear sigma model:
\begin{equation}
S={1\over{2\kappa^2_2}}\int d^2x\left[G_{ij}(X)\partial_{\mu}X^i
\partial^{\mu}X^i+\Lambda e^{W(X)}\right],
\label{sigact}
\end{equation}
where for the case of the two-dimensional model with the action 
(\ref{simp2dact})
\begin{equation}
(X^1,X^2)=(\phi,\rho),\ \ \  W(X)=2\rho+\ln V(\phi),\ \ \ 
\Lambda=1,\ \ \  G_{ij}=\left(\matrix{0&-1\cr -1&0}\right).
\label{sigactdef}
\end{equation}
The conditions for the non-linear sigma model action (\ref{sigact}) (with 
arbitrary $G_{ij}$ and $W$, not necessarily of the forms in Eq. 
(\ref{sigactdef})) to be invariant under a general variation $\delta X^i$ 
of the target space coordinates $X^i(x)$ are \cite{kst}
\begin{equation}
\delta X^i={{\epsilon^{ij}}\over{\sqrt{-|G|}}}W_{,j},\ \ \ \ \ 
\nabla_i\nabla_jW=0,
\label{confcond}
\end{equation}
where $|G|$ is the determinant of $G_{ij}$, $W_{,i}=\nabla_iW$ and 
$\nabla_i$ is the covariant derivative with respect to the 
target space metric $G_{ij}(X)$.  From the second condition in Eq. 
(\ref{confcond}), one can see that the two-dimensional target space (with 
the metric $G_{ij}$) of the general non-linear sigma model of the 
form (\ref{sigact}) has to be flat: $0=[\nabla_i,\nabla_j]W_{,k}=
{\cal R}_{\ell kij}W^{,\ell}$.  

Substituting the expression for $W(X)$ in Eq. (\ref{sigactdef}) into the 
second condition in Eq. (\ref{confcond}), one obtains the following 
condition \cite{cnn} for the classical conformal invariance of the general 
two-dimensional dilaton gravity model 
(\ref{simp2dact}):
\begin{equation}
{{d^2\ln V(\phi)}\over{d\phi^2}}=0,
\label{2dimgravconfcond}
\end{equation}
from which one obtains the following general expression for the potential 
$V(\phi)$ for the conformally invariant two-dimensional dilaton gravity 
models:
\begin{equation}
V(\phi)=4\lambda^2e^{\beta\phi},
\label{confpot}
\end{equation}
where $\lambda$ and $\beta$ are arbitrary constants.

Furthermore, the invariance condition (\ref{confcond}) implies the 
existence of a free field $F(X)$ defined by $F_{,k}=-\sqrt{-|G|}
\epsilon_{k\ell}W^{,\ell}$ \cite{kst}.  This free field $F$ is directly 
related to the Noether current $j_{\mu}$ associated with the symmetry under 
the general variation $\delta X^i$ as follows:
\begin{equation}
j_{\mu}=\partial_{\mu}X^iG_{ij}\delta X^j=\partial_{\mu}X^i\sqrt{-|G|}
\epsilon_{ij}W^{,j}=-\partial_{\mu}F.
\label{gennoethcurr}
\end{equation}
Making use of the symmetry invariance, one can put the general non-linear 
sigma model action (\ref{sigact}) (with arbitrary $G_{ij}$ and $W$) into 
the following form:
\begin{equation}
S={1\over{2\kappa^2_2}}\int d^2x{1\over\Upsilon}\left[-\partial_{\mu}F
\partial^{\mu}F+\partial_{\mu}W\partial^{\mu}W+\Lambda\Upsilon e^W\right],
\label{freelioubact}
\end{equation}
where $\Upsilon\equiv W^{,i}W_{,i}$ is a constant.  So, when it is invariant 
under the symmetry, the non-linear sigma model (\ref{sigact}) is described 
\cite {kst} by a free field $F(X)$ and a field $W(X)$ that satisfies the 
Liouville field equation in the flat spacetime:
\begin{equation}
\Box W={1\over 2}\Lambda\Upsilon e^W.
\label{wlibeq}
\end{equation}
We shall come back to this point later in this section.

In general, with an arbitrary potential $V(\phi)$ the two-dimensional 
dilaton gravity model with the action (\ref{simp2dact}) has the symmetry 
under the following transformations 
\cite{cnn}:
\begin{eqnarray}
\delta_E\phi&=&0,\ \ \ \ \ 
\delta_Eg_{\mu\nu}=g_{\mu\nu}a_{\rho}\nabla^{\rho}\phi
-{1\over 2}(a_{\mu}\nabla_{\nu}\phi+a_{\nu}\nabla_{\mu}\phi),
\cr
\delta_1\phi&=&0,\ \ \ \ \ 
\delta_1g_{\mu\nu}=\varepsilon_1\left[{{g_{\mu\nu}}\over{(\nabla\phi)^2}}
-2{{\nabla_{\mu}\phi\nabla_{\nu}\phi}\over{(\nabla\phi)^4}}\right],
\cr
\delta_2\phi&=&\varepsilon_2,\ \ \ \ \ 
\delta_2g_{\mu\nu}=\varepsilon_2V\left[{{g_{\mu\nu}}\over{(\nabla\phi)^2}}
-2{{\nabla_{\mu}\phi\nabla_{\nu}\phi}\over{(\nabla\phi)^4}}\right],
\cr
\delta_3\phi&=&0,\ \ \ \ \ 
\delta_3g_{\mu\nu}=-{{\varepsilon_3}\over 2}\left[g_{\mu\nu}+J
\left\{{{g_{\mu\nu}}\over{(\nabla\phi)^2}}-2{{\nabla_{\mu}\phi\nabla_{\nu}
\phi}\over{(\nabla\phi)^4}}\right\}\right],
\label{gen2dsym}
\end{eqnarray}
where $a_{\mu}$ is an arbitrary constant vector and $J^{\prime}(\phi)=
V(\phi)$.   The transformations $\delta_1$, $\delta_2$ and $\delta_3$ 
close down to a non-Abelian Lie algebra, in which $\delta_2$ is a central 
generator and $\delta_1$ and $\delta_3$ generate the affine 
subalgebra $[\delta_1,\delta_3]={1\over 2}\delta_1$.  
The Noether currents associated with the transformations in Eq. 
(\ref{gen2dsym}) are respectively
\begin{equation}
J^{\mu\nu}=Eg^{\mu\nu},\ \ \ 
j^{\mu}_1={{\nabla^{\mu}\phi}\over{(\nabla\phi)^2}},\ \ \ 
j^{\mu}_2=j^{\mu}_{\cal R}+V{{\nabla^{\mu}\phi}\over{(\nabla\phi)^2}},\ \ \ 
j^{\mu}_3=Ej^{\mu}_1,
\label{gen2dnoeth}
\end{equation}
where $E\equiv {1\over 2}\left[(\nabla\phi)^2-J(\phi)\right]$ is 
interpreted as the local energy of the configuration and 
$\nabla_{\mu}j^{\mu}_{\cal R}={\cal R}$.  The conservation of 
$J^{\mu\nu}$ implies that $E$ is a conserved scalar (independent of 
the spacetime coordinates), i.e. $\nabla_{\mu}E=0$.  When fields are 
on-shell, the symmetry $\delta_2$ is identified as a diffeomorphism 
with the vector field $f^{\mu}=\nabla^{\mu}\phi/(\nabla\phi)^2$.  
In particular, due to the conservation $\nabla_{\mu}E=0$ of the local 
energy $E$, the general action (\ref{simp2dact}) is invariant under the 
following transformation \cite{nav}:
\begin{eqnarray}
\delta_f\phi&=&0,
\cr
\delta_fg_{\mu\nu}&=&-\varepsilon f^{\prime}(E)\left[g_{\mu\nu}
-{{\nabla_{\mu}\phi\nabla_{\nu}\phi}\over{(\nabla\phi)^2}}\right]
+\varepsilon f(E)\left[{{g_{\mu\nu}}\over{(\nabla\phi)^2}}-2
{{\nabla_{\mu}\phi\nabla_{\nu}\phi}\over{(\nabla\phi)^4}}\right],
\label{genlocsym}
\end{eqnarray}
associated with the conserved Noether current $j^{\mu}_f=f(E)j^{\mu}_1$.  
The particular cases of $\delta_f$ with $f(E)=1,E$ respectively correspond 
to $\delta_1$ and $\delta_3$.  The transformation (\ref{genlocsym}) closes 
under the algebra $[\delta_f,\delta_g]={1\over 2}\delta_{f^{\prime}g-
g^{\prime}f}$ and in particular $\delta_f$'s with $f(E)=1,E,E^2$ close an 
$SL(2,{\bf R})$ algebra.  When the potential $V(\phi)$ takes the form 
(\ref{confpot}), the two-dimensional model (\ref{simp2dact}) is invariant 
under the conformal transformation, which is the linear combination 
$\delta_{\beta}=\delta_2+2\beta\delta_3$ of the symmetry transformations 
in Eq. (\ref{gen2dsym}) \cite{cnn}:
\begin{equation}
\delta_{\beta}g_{\mu\nu}=-\beta\varepsilon g_{\mu\nu},\ \ \ \ \ 
\delta_{\beta}\phi=\varepsilon,
\label{genconfsym}
\end{equation}
where keep in mind that the bars are suppressed in the above and we have let 
$\varepsilon:=\varepsilon_2=\varepsilon_3$.

\subsection{Dilatonic 0-brane case}

From the near horizon metric (\ref{nearhordualmet}) or (\ref{horomet}) 
for the dual-frame dilatonic 0-brane solution, one can see that there is 
the Freund-Rubin compactification \cite{fre} on $S^{D-2}$ of the action 
(\ref{ddim0actdual2}) down to the following two-dimensional effective gauged 
supergravity action \cite{bbh}:
\begin{equation}
S={1\over{2\kappa^2_2}}\int d^2x\sqrt{-g}e^{\delta\phi}
\left[{\cal R}_g+\gamma(\partial\phi)^2+\Lambda\right],
\label{2dact}
\end{equation}
where the parameters in this action are defined as
\begin{eqnarray}
\delta&\equiv&-{{(D-2)a}\over{D-3}},\ \ \ \ \ \ 
\gamma\equiv {{D-1}\over{D-2}}\delta^2-{4\over{D-2}},
\cr 
\Lambda&\equiv&{{(D-3)}\over{2\mu^2}}\left[2(D-2)-{{4(D-3)}\over\Delta}
\right].
\label{2dactpara}
\end{eqnarray}

To bring the action (\ref{2dact}) to the standard form (\ref{gen2dact}), 
one redefines the dilaton as $\Phi=e^{\delta\phi}$ and then applies the 
Weyl rescaling of the metric $g_{\mu\nu}=\Phi^{-{\gamma\over{\delta^2}}}
e^{\Phi\over 2}\tilde{g}_{\mu\nu}$.  Then, the action (\ref{2dact}) takes 
the following standard form:
\begin{equation}
S={1\over{2\kappa^2_2}}\int d^2x\sqrt{-\tilde{g}}\left[\Phi
{\cal R}_{\tilde{g}}+{1\over 2}\partial_{\mu}\Phi\partial^{\mu}\Phi
+\Lambda\Phi^{1-{\gamma\over{\delta^2}}}e^{\Phi\over 2}\right].
\label{stnddilact}
\end{equation}
To remove the kinetic term for $\Phi$ to bring the action 
(\ref{stnddilact}) to the form (\ref{simp2dact}), one applies one more 
Weyl rescaling transformation $\tilde{g}_{\mu\nu}=e^{-{\Phi\over 2}}
\bar{g}_{\mu\nu}$.  The resulting action has the following form:
\begin{equation}
S={1\over{2\kappa^2_2}}\int d^2x\sqrt{-\bar{g}}\left[\Phi
{\cal R}_{\bar{g}}+\Phi^{1-{{\gamma}\over{\delta^2}}}\Lambda\right].
\label{2dactonepot}
\end{equation}

This resulting two-dimensional dilaton gravity model has the $SL(2,{\bf R})$ 
affine symmetry, as will be discussed in the next subsection.  This 
$SL(2,{\bf R})$ symmetry is expected from the fact that the generalized 
conformal mechanics of the probe dilatonic 0-brane also has the 
$SL(2,{\bf R})$ symmetry.  Since the potential term in the action 
(\ref{2dactonepot}) is not of the form (\ref{confpot}), the two-dimensional 
gravity model (\ref{2dactonepot}), derived from the dilatonic 0-brane 
supergravity action in $D$ dimensions, is not conformal in general.  This is 
also consistent with the fact that the boundary theory counterpart to the 
bulk near-horizon dilatonic 0-brane theory, namely the generalized conformal 
quantum mechanics of the probe dilatonic 0-brane, is not genuinely conformal.  
From these facts, we are lead to the speculation that the generalized 
conformal quantum mechanics of the probe dilatonic 0-brane is dual to the 
two-dimensional dilaton gravity model with the classical action given by Eq. 
(\ref{stnddilact}) or Eq. (\ref{2dactonepot}).  In particular for the 
D0-brane case, which is relevant to the M-theory in the light-cone frame 
\cite{sus3}, the exponent in the potential is given by $1-\gamma/\delta^2=5/9$.

The model with the action (\ref{2dactonepot}) is conformal when $\gamma=
\delta^2$.  From the expressions for $\gamma$ and $\delta$ in Eq. 
(\ref{2dactpara}), one can see that this happens when the dilaton coupling 
parameter $a$ takes the following special value:
\begin{equation}
a=2{{D-3}\over{D-2}}.
\label{dilcoupconf}
\end{equation}
A special case of interest is the $D=4$ case.  In this case, the dilaton 
coupling parameter becomes $a=1$, which corresponds to the string theory 
inspired model of the $D=4$ Einstein-Maxwell-dilaton theory.  Generally, 
for any values of $D$ with $a$ taking the values specified by Eq. 
(\ref{dilcoupconf}), one can bring the action (\ref{2dactonepot}) to the 
form of the action of the Callan, Giddings, Harvey and Strominger (CGHS) 
model \cite{cghs}.  This can be done by first redefining the dilaton field as 
$\Phi=e^{-2\phi}$ and then by applying the Weyl rescaling transformation 
$g_{\mu\nu}=e^{2\phi}\bar{g}_{\mu\nu}$.  The resulting action has the 
following form:
\begin{equation}
S={1\over{2\kappa^2_2}}\int d^2x\sqrt{-g}e^{-2\phi}\left[{\cal R}_g
+4g^{\mu\nu}\partial_{\mu}\phi\partial_{\nu}\phi+\Lambda\right].
\label{dilcghsact}
\end{equation}
Indeed, the CGHS model is known to be conformally invariant.  Namely, 
the CGHS model is invariant under the following transformation: 
\begin{equation}
\delta\phi=\varepsilon e^{2\phi},\ \ \ \ \ 
\delta g_{\mu\nu}=2\varepsilon e^{2\phi}g_{\mu\nu}.
\label{CGHSconf}
\end{equation}
This transformation, when expressed in terms of the fields of the action 
(\ref{2dactonepot}) with $\gamma=\delta^2$, corresponds to the linear 
combination \cite{cnn} $\delta=\delta_2-2\lambda^2\delta_1$ of the 
symmetry transformations (\ref{gen2dsym}) of the generic action 
(\ref{simp2dact}) with the potential given by Eq. (\ref{confpot}) with 
$\beta=0$:
\begin{equation}
\delta\Phi=\varepsilon,\ \ \ \ \ 
\delta\bar{g}_{\mu\nu}=0.
\label{CGHStran}
\end{equation} 

Another interesting case is when $\gamma=0$, in which case the action 
(\ref{2dactonepot}) describes the Jackiw-Teitelboim \cite{jac,tei} model with 
the following action:
\begin{equation}
S={1\over{2\kappa^2_2}}\int d^2x\sqrt{-\bar{g}}\left[\Phi
{\cal R}_{\bar{g}}+\Phi\Lambda\right].    
\label{JTmodel}
\end{equation}
This happens when the dilaton coupling $a$ takes the following form:
\begin{equation}
a={2\over\sqrt{D-1}}{{D-3}\over{D-2}}.
\label{JTdilcoup}
\end{equation}
An example is the case of $D=4$ dilatonic 0-brane with the dilaton 
coupling $a=1/\sqrt{3}$.

\subsection{Non-dilatonic 0-brane case}

The non-dilatonic case (i.e. $a=0$ and $\phi=0$) requires special 
treatment, because the general two-dimensional action (\ref{2dact}) 
(as well as the two-dimensional actions (\ref{stnddilact}) and 
(\ref{2dactonepot}) in standard forms) is not well-defined for this 
case, behaving singularly.  For example, the general formula for the 
cosmological constant $\Lambda$ in Eq. (\ref{2dactpara}) becomes zero 
when $a=0$, although the compactified $D=2$ solution in the near-horizon 
region is the AdS$_2$ space, which requires non-zero cosmological constant 
term in the action.  So, in the following we first derive the compactified 
two-dimensional action for the non-dilatonic 0-brane case separately.  

The effective action for the non-dilatonic 0-brane in $D$ spacetime dimensions 
is 
\begin{equation}
S={1\over{2\kappa^2_D}}\int d^Dx\sqrt{-G}\left[{\cal R}_{G}
-{1\over{2\cdot 2!}}F_{MN}F^{MN}\right],
\label{nondilddim0act}
\end{equation}
where ${\cal R}_{G}$ is the Ricci scalar of the $D$-dimensional spacetime 
metric $G_{MN}$ ($M,N=0,1,...,D-1$) and $F_{MN}$ is the field strength of 
the $U(1)$ gauge field $A_M$.  We take the following Ansatz for the 
$D$-dimensional spacetime metric:
\begin{equation}
G_{MN}dx^Mdx^N=g_{\mu\nu}(x^{\mu})dx^{\mu}dx^{\nu}+
\exp\left[-{4\over{D-2}}\sigma(x^{\mu})\right]d\Omega^2_{D-2},
\label{ddimmet}
\end{equation}
where $\mu,\nu=0,1$.  This general metric Ansatz includes the metric for 
the 0-brane solution as a special case.  We assume that the $U(1)$ gauge 
field $A_M$ is only electric, i.e. its only non-zero component is $A_t$.  
Then, by solving the Maxwell's equation $\nabla_{M}F^{MN}=0$ with the 
above metric Ansatz (\ref{ddimmet}), one obtains the following expression 
for the electric field $E=F_{tr}$: 
\begin{equation}
F_{tr}=Qe^{2\sigma}\sqrt{-g},
\label{emfld}
\end{equation} 
where $Q$ is a constant related to the electric charge and 
$g\equiv\det(g_{\mu\nu})$.  Then, the $D$-dimensional action 
(\ref{nondilddim0act}), upon dimensional reduction on $S^{D-2}$, becomes 
of the following form:
\begin{eqnarray}
S&=&{1\over{2\kappa^2_2}}\int d^2x\sqrt{-g}e^{-2\sigma}\left[{\cal R}_g
-4{{D-3}\over{D-2}}\partial_{\mu}\sigma\partial^{\mu}\sigma
-(D-2)(D-3)e^{{4\over{D-2}}\sigma}+{{Q^2}\over 2}e^{4\sigma}\right],
\cr
&=&{1\over{2\kappa^2_2}}\int d^2x\sqrt{-g}\left[\bar{\sigma}{\cal R}_g
-{{D-3}\over{D-2}}{{\partial_{\mu}\bar{\sigma}\partial^{\mu}\bar{\sigma}}
\over{\bar{\sigma}}}-(D-2)(D-3)\bar{\sigma}^{{D-4}\over{D-2}}+{{Q^2}\over 2}
\bar{\sigma}^{-1}\right],
\label{2dacteinmax}
\end{eqnarray}
where ${\cal R}_g$ is the Ricci scalar for the two-dimensional metric 
$g_{\mu\nu}$ defined in Eq. (\ref{ddimmet}) and $\bar{\sigma}\equiv 
e^{-2\sigma}$.  The field equations that follow from this action are
\begin{eqnarray}
& &{\cal R}_g+{{D-3}\over{D-2}}\left[{{\partial_{\mu}\bar{\sigma}
\partial^{\mu}\bar{\sigma}}\over{\bar{\sigma}^2}}-2\nabla_{\mu}
\left({{\partial^{\mu}\bar{\sigma}}\over{\bar{\sigma}}}\right)\right]
-(D-3)(D-4)\bar{\sigma}^{-{2\over{D-2}}}-{{Q^2}\over 2}\bar{\sigma}^{-2}=0,
\cr
& &\nabla_{\mu}\nabla_{\nu}\bar{\sigma}+{{D-3}\over{D-2}}\left[{1\over 2}
g_{\mu\nu}{{\partial_{\rho}\bar{\sigma}\partial^{\rho}\bar{\sigma}}\over
{\bar{\sigma}}}-{{\partial_{\mu}\bar{\sigma}\partial_{\nu}\bar{\sigma}}
\over{\bar{\sigma}}}\right]
\cr
& &\ \ \ \ \ \ \ \ \ \ \ \ \ \ \ \ \ \ \ \ \ \ \ \ \ \ \ \ \ \ \ \ \ \ \ \ \ \ 
+{1\over 2}g_{\mu\nu}\left[(D-2)(D-3)\bar{\sigma}^{{D-4}\over{D-2}}
-{{Q^2}\over 2}\bar{\sigma}^{-1}\right]=0,
\label{2dimeqnmtn}
\end{eqnarray}
where $\nabla_{\mu}$ is the covariant derivative with respect to the 
metric $g_{\mu\nu}$.  
The solution to these field equations, through the relations (\ref{ddimmet}) 
and (\ref{emfld}) between the 2-dimensional fields and the $D$-dimensional 
fields, reproduces non-dilatonic 0-brane solution in $D$ dimensions. 

Note, in this paper we are particularly interested in the two-dimensional 
theory associated with the near-horizon approximation of the 0-brane solutions.
The solution for the non-dilatonic 0-brane in $D$ dimensions is
\begin{equation}
ds^2=-H^{-2}dt^2+H^{2\over{D-3}}(dr^2+r^2d\Omega^2_{D-2}),\ \ \ \ 
A_t=-H^{-1},
\label{nondil0branesol}
\end{equation}
where $H=1+\left({{\mu}\over{r}}\right)^{D-3}$.  In the near-horizon 
region ($r\ll\mu$), the solution takes the following form:
\begin{equation}
ds^2=-\left({{r}\over{\mu}}\right)^{2(D-3)}dt^2+\left({{\mu}\over{r}}
\right)^2dr^2+\mu^2d\Omega^2_{D-2},\ \ \ \ \ 
A_t=-\left({{r}\over{\mu}}\right)^{D-3}.
\label{nondilnearhor}
\end{equation}
As expected, the spacetime in the near-horizon region is AdS$_2\times 
S^{D-2}$.  So, the two-dimensional scalar field $\sigma(x^{\mu})$ (or 
$\bar{\sigma}(x^{\mu})$) defined in Eq. (\ref{ddimmet}) becomes constant 
in the near-horizon region of the 0-brane solution.  
The field equations (\ref{2dimeqnmtn}) therefore reduce to the following:
\begin{equation}
{\cal R}_g-\Lambda=0,
\label{simpfldeq}
\end{equation}
where $\Lambda=2\left({{D-3}\over\mu}\right)^2$.  This constant $\Lambda$ 
is exactly the curvature ${\cal R}_g$ of the $(t,r)$ part $g_{\mu\nu}
dx^{\mu}dx^{\nu}=-\left({{r}\over{\mu}}\right)^{2(D-3)}dt^2+\left({{\mu}
\over{r}}\right)^2dr^2$ of the metric $G_{MN}$ in Eq. (\ref{nondilnearhor}).  
Note, the curvature ${\cal R}_g$ of this AdS$_2$ metric has the 
positive sign due to the choice of the signature $(-+)$ for the metric.  

We realize that the field equation (\ref{simpfldeq}) describes the Liouville 
theory (for a review on the Liouville theory, see Ref. \cite{sei}).  
This can be seen by going to the coordinate system where the spacetime 
becomes conformally flat, which is always possible for the two-dimensional 
spacetime.  For the two-dimensional subset of the spacetime described by 
the metric in Eq. (\ref{nondilnearhor}), this is achieved by 
defining new radial coordinate as $\rho\equiv{{\mu}\over{D-3}}\left({{\mu}
\over{r}}\right)^{D-3}$.  In this new coordinates, the solution 
(\ref{nondilnearhor}) takes the following form:
\begin{equation}
ds^2=\left({{\mu}\over{D-3}}\right)^2{1\over{\rho^2}}\left[-dt^2+
d\rho^2\right]+\mu^2d\Omega^2_{D-2},\ \ \ \ \ 
A_t=-{{\mu}\over{D-3}}{1\over{\rho}},
\label{conflatmet}
\end{equation}
where the $(t,\rho)$ part of the metric is conformally flat.
If we denote the two-dimensional (subspace) metric $g_{\mu\nu}$ in the 
conformally flat coordinate system as $g_{\mu\nu}=e^{\varphi}\eta_{\mu\nu}$, 
then Eq. (\ref{simpfldeq}) reduces to the following:
\begin{equation}
\eta^{\mu\nu}\partial_{\mu}\partial_{\nu}\varphi-\Lambda 
e^{\varphi}=0,
\label{lioueqn}
\end{equation}
which is the Liouville field equation in flat spacetime.

In general, the Liouville theory in a curved spacetime with the metric 
$\hat{g}_{\mu\nu}$ is defined by the following action \cite{pol1}:
\begin{equation}
S={1\over {8\pi}}\int d^2x\sqrt{-\hat{g}}\left[\hat{g}^{\mu\nu}\partial_{\mu}
\phi\partial_{\nu}\phi+Q\phi {\cal R}_{\hat{g}}+{{2\Lambda}\over
{\gamma^2}}e^{\gamma\phi}\right].
\label{lioucurv}
\end{equation}
The requirement of conformal invariance of the classical 
Liouville action determines the classical background charge 
coefficient $Q$ to be $Q=2/\gamma$.  With this choice of $Q$, the 
Liouville action (\ref{lioucurv}) is invariant under the following 
Weyl rescaling transformations:
\begin{equation}
\hat{g}_{\mu\nu}\to e^{2\sigma}\hat{g}_{\mu\nu},\ \ \ \ \ 
\gamma\phi\to \gamma\phi-2\sigma,
\label{liouweyltran}
\end{equation}
which is equivalent to demanding the metric $g_{\mu\nu}=e^{\gamma\phi}
\hat{g}_{\mu\nu}$ to be invariant.  

The Liouville theory is known to have the hidden $SL(2,{\bf R})$ symmetry 
\cite{pol2,pol3}.  This hidden $SL(2,{\bf R})$ symmetry can be seen
\footnote{Before one applies the analysis described in the following, 
one has to first apply the Weyl rescaling transformation $\hat{g}_{\mu\nu}\to
e^{\gamma\phi}\hat{g}_{\mu\nu}$ to remove the dependence on $\phi$ of 
the potential term in the action (\ref{lioucurv}).} 
easily in the light-cone (Polyakov) gauge, in which 
the metric takes the following form:
\begin{equation}
ds^2=dx^+dx^-+h_{++}(x^+,x^-)(dx^+)^2.
\label{lighconguag}
\end{equation}
The residual symmetry of this gauge choice contains the Virasoro (conformal) 
symmetry and the $SL(2,{\bf R})$ current symmetry of the Liouville theory.  
One can replace the metric component $h_{++}$ by a new field $f$ through the 
relation:
\begin{equation}
\partial_+f=h_{++}\partial_-f.
\label{metcomptofld}
\end{equation}
By using the equation of motion for $\phi$, one can also express the 
Liouville field $\phi$ in terms of $f$.  The resulting Liouville action, 
expressed totally in terms of $f$, is invariant under the following 
variation of $f$, which corresponds to the reparametrization variation
\footnote{When applying the reparametrization variation, one has to 
make sure that the light-cone gauge is maintained.} 
$\delta h_{++}=
\nabla_+\varepsilon^-$ of the metric component $h_{++}$:
\begin{equation}
\delta f=\varepsilon^-\partial_-f,
\label{hppvar}
\end{equation}
provided the infinitesimal parameter $\varepsilon^-(x)$ satisfies the 
condition $\partial^3_-\varepsilon^-=0$.  The most general infinitesimal 
variation parameter $\varepsilon^-$ that satisfies this condition is
\begin{equation}
\varepsilon^-(x^+,x^-)=w_-(x^+)+x^-w_0(x^+)+(x^-)^2w_+(x^+).
\label{genvarforf}
\end{equation}
Under this reparametrization variation with the infinitesimal parameter 
given by (\ref{genvarforf}), the metric component $h_{++}$ transforms as
\begin{equation}
\delta h_{++}=\left[w_-j^-+w_0j^0+w_+j^+\right]h_{++}+
2\left[\partial_+w_-+x^-\partial_+w_0+(x^-)^2\partial_+w_+\right],
\label{metvar}
\end{equation}
where $j^-\equiv\partial_-$, $j^0\equiv x^-\partial_--1$ and 
$j^+\equiv(x^-)^2\partial_--2x^-$.  The generators $j^{-,0,+}$ of 
the above transformation satisfy the following $SL(2,{\bf R})$ algebra:
\begin{equation}
[j^0,j^-]=-j^-,\ \ \ \  [j^0,j^+]=j^+,\ \ \ \  [j^+,j^-]=-2j^0.
\label{libcrvsl2r}
\end{equation}
Another way of seeing the hidden $SL(2,{\bf R})$ is by considering the 
metric component constraint $\partial^3_-h_{++}=0$ that follows from the 
equations of motion.  The general form of the metric component $h_{++}$ 
that satisfies this constraint is
\begin{equation}
h_{++}(x^+,x^-)=J^+(x^+)-2J^0(x^+)x^-+J^{-}(x^+)(x^-)^2.
\label{genconmet}
\end{equation}
Under the reparametrization variation $\delta h_{++}=\nabla_+\varepsilon^-$ 
with the infinitesimal parameter $\epsilon^-$ given by Eq. (\ref{genvarforf}), 
$J^a(x^+)$ ($a=-,0,+$) transform as
\begin{equation}
\delta J^a=f^{abc}w_bg_{cd}J^d+2g^{ab}\partial_+w_b,
\label{jsl2tran}
\end{equation}
where $f^{abc}$ and $g_{ab}$ are respectively the structure constants and 
the Cartan's metric of the $SL(2,{\bf R})$ algebra.  This $SL(2,{\bf R})$ 
affine Kac-Moody symmetry can also be seen by considering the Ward identity 
\cite{pol2}.  

Similarly, one can show that the generic two-dimensional dilaton 
gravity action (\ref{stnddilact}) obtained from the (source) dilatonic 
0-brane action in $D$ dimensions also has the $SL(2,{\bf R})$ current 
symmetry.  So, the $SL(2,{\bf R})$ symmetry of the (generalized) conformal 
quantum mechanics of the (probe) dilatonic and non-dilatonic 0-branes can 
be realized within the two-dimensional dilaton gravity models obtained from 
the effective actions for the (source) 0-branes through the $S^{D-2}$ 
compactification.  Before one applies the analysis similar to the one 
described in the previous paragraph, one has to first apply the Weyl rescaling 
transformation of the metric to remove the dilaton dependence of the 
potential term of the action (\ref{stnddilact}) and then redefine the 
dilaton to have the standard dilaton kinetic term.  However, unlike the 
case of the two-dimensional dilaton gravity model associate with the (source) 
dilatonic 0-branes (discussed in the previous subsection), the Liouville 
theory, associated with the (source) non-dilatonic 0-branes, in addition 
has the conformal symmetry, which we discuss in the following.   

The stress-energy tensor $T_{\mu\nu}=2\pi\delta S/\delta\hat{g}^{\mu\nu}$ of 
the classical Liouville theory with the action (\ref{lioucurv}) has the 
following form:
\begin{equation}
T_{z\bar{z}}=0,\ \ \ \ \ 
T_{zz}=-{1\over 2}(\partial\phi)^2+{1\over 2}Q\partial^2\phi\equiv T(z),
\ \ \ \ \ 
T_{\bar{z}\bar{z}}=\bar{T}(\bar{z}),
\label{liuvtens}
\end{equation}
where $z=t+ix$.  Since the trace of the stress-energy tensor 
$T^{\mu}_{\ \mu}=4T_{z\bar{z}}$ is zero, the classical Liouville theory is 
conformally invariant, which is consistent with the fact that the conformal 
quantum mechanics of non-dilatonic 0-branes also has conformal symmetry.  
Under the conformal transformation $z\to w(z)$, the Liouville field and 
the stress-energy tensor transform as
\begin{equation}
\phi\to\phi-{1\over\gamma}\ln\left|{{dw}\over{dz}}\right|^2,\ \ \ 
T(z)\to\left({{dw}\over{dz}}\right)^2T_{ww}+{1\over{\gamma^2}}S[w;z],
\label{conftrnliv}
\end{equation}
where $S[w;z]\equiv {{w^{\prime\prime\prime}}\over{w^{\prime}}}-{3\over 2}
\left({{w^{\prime\prime}}\over{w^{\prime}}}\right)^2$ 
is the Schwartzian derivative.  Therefore, the coefficients (called Virasoro 
operators) of the following Laurent expansion of the stress-energy tensor 
\begin{equation}
T(z)=\sum_{n\in{\bf Z}}z^{-n-2}L_n,\ \ \ \ \ 
\bar{T}(\bar{z})\sum_{n\in{\bf Z}}\bar{z}^{-n-2}\bar{L}_n,
\label{laurent}
\end{equation}
where $L_n=\oint{{dz}\over{2\pi i}}z^{n+1}T(z)$ and 
$\bar{L}_n=\oint{{d\bar{z}}\over{2\pi i}}\bar{z}^{n+1}\bar{T}(\bar{z})$, 
satisfy the following Virasoro algebra with the central charge 
$c=12/\gamma^2$:
\begin{eqnarray}
[L_n,L_m]&=&(n-m)L_{n+m}+{1\over{\gamma^2}}(n^3-n)\delta_{n+m,0},
\cr
[\bar{L}_n,\bar{L}_m]&=&(n-m)\bar{L}_{n+m}+{1\over{\gamma^2}}(n^3-n)
\delta_{n+m,0},
\cr
[L_n,\bar{L}_m]&=&0.
\label{viraalg}
\end{eqnarray}
Note, the conformal algebra of the Liouville theory already has the anomaly 
term (proportional to the central charge) at the classical level.  The 
quantum effect due to the normal ordering of the Virasoro generators 
gives rise to the additional contribution to the central charge.  This 
results in the renormalization of the parameter $1/\gamma^2$ in the Liouville 
action (\ref{lioucurv}) \cite{gn}.  

In the flat background ($\hat{g}_{\mu\nu}=\eta_{\mu\nu}$), the Liouville 
action (\ref{lioucurv}) takes the following form, after the field 
redefinition $\phi=\gamma^{-1}\varphi$:
\begin{equation}
S={1\over{4\pi\gamma^2}}\int dx^2\left[{1\over 2}\partial_{\mu}\varphi
\partial^{\mu}\varphi+\Lambda e^{\varphi}\right].
\label{liouflat}
\end{equation}
The field equation of this action yields the Liouville field equation 
(\ref{lioueqn}) in flat spacetime, which is also the field equation of 
the non-dilatonic 0-brane in the near-horizon region.  This flat spacetime 
Liouville action is invariant under the following conformal transformation:
\begin{eqnarray}
x^{\pm}&\to& f^{\pm}(x^{\pm}),
\cr
\varphi(x^+,x^-)&\to&\varphi(f^+(x^+),f^-(x^-))+\ln\left[f^{+\,\prime}(x^+)
f^{-\,\prime}(x^-)\right],
\label{2dconftranmet}
\end{eqnarray}
where $x^{\pm}=t\pm x$ and $f^{\pm\,\prime}=df^{\pm}/dx^{\pm}$.  Again, 
this is consistent with the fact that the boundary theory of the bulk theory 
on the AdS$_2$ space, i.e. the quantum mechanics of the (probe) non-dilatonic 
0-brane, is also conformal.  

The general solution to the Liouville equation (\ref{lioueqn}) in flat 
spacetime is given by \cite{liv}
\begin{equation}
\varphi(x^+,x^-)=\ln{{8A^{\prime}_+(x^+)A^{\prime}_-(x^-)}\over
{|\Lambda|\left[A_+(x^+)-\epsilon A_-(x^-)\right]^2}},
\label{liougensol}
\end{equation}
where $A^{\pm}(x^{\pm})$ is an arbitrary function such that 
$A^{\prime}_{\pm}=dA_{\pm}/dx^{\pm}>0$ and $\epsilon$ is 
the sign of $\Lambda$, i.e. $\epsilon=\Lambda/|\Lambda|$.  This 
general solution is invariant under the following $SL(2,{\bf R})$ 
transformation of $A_{\pm}$:
\begin{equation}
A_+(x^+)\to {{aA_+(x^+)+b}\over{cA_+(x^+)+d}},\ \ \ \ \ 
A_-(x^-)\to {{aA_-(x^-)+\epsilon b}\over{\epsilon cA_-(x^-)+d}},
\label{bianchtrn}
\end{equation}
where real numbers $a,b,c,d$ satisfy $ad-bc=1$.  

The underlying $SL(2,{\bf R})$ symmetry of the Liouville theory in flat
spacetime becomes manifest in the Hamiltonian expressed in terms of free 
fields and their conjugate momenta.  Such free fields are obtained 
from $A_+(x^+)$ through the `inverse scattering method' followed by the 
`bosonization'.  As a side comment, the fields $\psi_{\pm 1/2}$ obtained 
from $A_+(x^+)$ through the inverse scattering method transform in the 
linear spin 1/2 representation under $SL(2,{\bf R})$.  The details on 
the hidden $SL(2,{\bf R})$ symmetry of the Liouville theory in flat spacetime 
can be found, for example, in Ref. \cite{sei}.

The general solution (\ref{liougensol}) can be obtained from the following 
simple solution:
\begin{equation}
\varphi(x^+,x^-)=\ln{{8}\over{|\Lambda|(x^+-\epsilon x^-)^2}},
\label{simpliousol}
\end{equation}
which corresponds to the $A_{\pm}=x^{\pm}$ case of (\ref{liougensol}), by 
applying the conformal transformation (\ref{2dconftranmet}).  The resulting 
(conformal transformed) general solution is given by Eq. (\ref{liougensol}) 
with $A_{\pm}(x^{\pm})=f^{\pm}(x^{\pm})$.  In fact, under the conformal 
transformation (\ref{2dconftranmet}), the functions $A_{\pm}(x^{\pm})$ in 
the general solution (\ref{liougensol}) transform as in Eq. 
(\ref{bianchtrn}), namely
\begin{equation}
A_+(f^+(x^+))={{aA_+(x^+)+b}\over{cA_+(x^+)+d}},\ \ \ \ \ 
A_-(f^-(x^-))={{aA_-(x^-)+\epsilon b}\over{\epsilon cA_-(x^-)+d}}.
\label{confbianchtrn}
\end{equation}
Therefore, all the solutions to the Liouville equation (\ref{lioueqn}) in 
the flat spacetime is related to the simple solution (\ref{simpliousol}) 
through the conformal transformation (\ref{2dconftranmet}) or the 
$SL(2,{\bf R})$ transformation (\ref{bianchtrn}).  

For the case under consideration in this section, namely the near horizon 
spacetime of the non-dilatonic 0-brane, the constant $\Lambda=
2\left({{D-3}\over{\mu}}\right)^2$ is always positive and therefore in 
the above $\epsilon=1$.  We notice that the simplest solution 
(\ref{simpliousol}) to the Liouville equation (\ref{lioueqn}) corresponds 
to the two-dimensional subspace of the near-horizon solution 
(\ref{conflatmet}) of the non-dilatonic 0-brane in the conformal 
coordinates.  Therefore, all the solutions of the Liouville field theory is 
locally related through the conformal transformation (\ref{2dconftranmet}) 
or the $SL(2,{\bf R})$ transformation (\ref{bianchtrn}) to the near-horizon 
region solution of the non-dilatonic 0-brane.  This provides one of the 
evidences for the equivalence of the conformal quantum mechanics of 
non-dilatonic 0-branes to the Liouville field theory.  

We mentioned previously that the most general two-dimensional dilaton 
gravity model with conformal invariance has the following action:
\begin{equation}
S={1\over{2\kappa^2_2}}\int d^2x\sqrt{-\bar{g}}\left[\bar{\phi}
{\cal R}_{\bar{g}}+4\lambda^2e^{\beta\bar{\phi}}\right].
\label{conf2dact}
\end{equation}
For the $\beta\neq 0$ case, by redefining a new dilaton field as $\varphi
\equiv 2\beta\bar{\phi}$ and then by applying the Weyl rescaling 
transformation $g_{\mu\nu}=e^{-\varphi/2}\bar{g}_{\mu\nu}$, one can 
transform the action (\ref{conf2dact}) to the following form:
\begin{equation}
S={1\over{4\beta\kappa^2_2}}\int d^2x\sqrt{-g}\left[\varphi{\cal R}_g+
{1\over 2}g^{\mu\nu}\partial_{\mu}\varphi\partial_{\nu}\varphi+
8\beta\lambda^2e^{\varphi}\right].
\label{transfconf2act}
\end{equation}
We recognize that this is the Liouville action in curved spacetime.  In the 
$\beta=0$ case, one can also bring the action (\ref{conf2dact}) to the 
Liouville-like form through the Weyl rescaling transformation $g_{\mu\nu}=
e^{\bar{\phi}}\bar{g}_{\mu\nu}$. The resulting action has the following form:
\begin{equation}
S={1\over{2\kappa^2_2}}\int d^2x\sqrt{-g}\left[\bar{\phi}{\cal R}_g+
g^{\mu\nu}\partial_{\mu}\bar{\phi}\partial_{\nu}\bar{\phi}+
4\lambda^2e^{\bar{\phi}}\right],
\label{tranfconf2act}
\end{equation}
with different numerical factor in front of the kinetic term of 
$\bar{\phi}$, and therefore has to be identified with the Liouville 
action (\ref{lioucurv}) with non-critical value of $Q$ different from 
$Q=2/\gamma$.  Anyhow, by going to the gauge and the coordinate frame 
in which the spacetime is flat, one can bring the actions 
(\ref{transfconf2act}) and (\ref{tranfconf2act}) to the form of the 
action (\ref{liouflat}) for the Liouville theory in flat spacetime.  
This can also be seen directly from the following field equations 
for the action (\ref{conf2dact}) in the conformal coordinates (in which 
the metric takes the form $g_{\mu\nu}=e^{2\rho}\eta_{\mu\nu}$):
\begin{eqnarray}
\partial_{\mu}\partial^{\mu}(2\rho-\beta\bar{\phi})&=&0
\cr
\partial_{\mu}\partial^{\mu}(2\rho+\beta\bar{\phi})&=&8\lambda^2\beta
e^{\rho+\beta\bar{\phi}}.
\label{expmodeqn}
\end{eqnarray}
So, whereas $2\rho-\beta\bar{\phi}$ is a free field, 
$2\rho+\beta\bar{\phi}$ satisfies the Liouville field equation in 
flat spacetime.  If one lets $2\rho=\beta\bar{\phi}$, which is 
equivalent to applying the Weyl rescaling transformation $g_{\mu\nu}=
e^{-\beta\bar{\phi}}\bar{g}_{\mu\nu}$ and then going to the gauge 
where $g_{\mu\nu}=\eta_{\mu\nu}$, one is left only with a theory with the 
Liouville field $\varphi=2\rho+\beta\bar{\phi}=2\beta\bar{\phi}$ in flat 
spacetime, as pointed out in the above.  Also, as it is pointed out in 
the paragraph that follows Eq. (\ref{confpot}), the general non-linear 
sigma model with the action (\ref{sigact}) with the conformal symmetry 
is described by a free scalar field and the (flat spacetime) Liouville 
field.  Thus, one can see that all the conformally invariant two-dimensional 
dilaton gravity is at least locally equivalent to the Liouville theory in 
flat spacetime and therefore to the conformal quantum mechanics of the probe 
non-dilatonic 0-brane in the near-horizon background of the source 
non-dilatonic 0-brane.

\end{document}